\documentclass{aa}  

\usepackage{graphicx}
\usepackage{txfonts}
\usepackage[colorlinks,citecolor=blue]{hyperref}
\begin{document} 

\title{Extreme ion acceleration at extragalactic jet termination shocks}

\titlerunning{Extreme ion acceleration at extragalactic jet termination shocks}

\author{Beno\^it Cerutti \inst{1}\and Gwenael Giacinti\inst{2,3}}

\institute{Univ. Grenoble Alpes, CNRS, IPAG, 38000 Grenoble, France\\
           \email{benoit.cerutti@univ-grenoble-alpes.fr}
           \and
           Tsung-Dao Lee Institute, Shanghai Jiao Tong University, Shanghai 201210, P. R. China
           \and
           School of Physics and Astronomy, Shanghai Jiao Tong University, Shanghai 200240, P. R. China\\
           \email{gwenael.giacinti@sjtu.edu.cn}
           }

\date{Received \today; accepted \today}

 
\abstract
{Extragalactic plasma jets are some of the few astrophysical environments able to confine ultra-high-energy cosmic rays, but whether they are capable of accelerating these particles is unknown.}
{In this work, we revisit particle acceleration at relativistic magnetized shocks beyond the local uniform field approximation by considering the global transverse structure of the jet.}
{Using large two-dimensional particle-in-cell simulations of a relativistic electron-ion plasma jet, we show that the termination shock forming at the interface with the ambient medium accelerates particles up to the confinement limit.}
{The radial structure of the jet magnetic field leads to a relativistic velocity shear that excites a von~K\'arm\'an vortex street in the downstream medium trailing behind an over-pressured bubble filled with cosmic rays. Particles are efficiently accelerated at each crossing of the shear flow boundary layers.}
{These findings support the idea that extragalactic plasma jets may be capable of producing ultra-high-energy cosmic rays. This extreme particle acceleration mechanism may also apply to microquasar jets.}

\keywords{acceleration of particles -- shock waves -- galaxies: jets -- methods: numerical}
               
\maketitle


\section{Introduction}

Active supermassive black holes nested in the heart of galaxies produce powerful plasma outflows in the form of relativistic magnetized jets \citep{2019ARA&A..57..467B, 2020NewAR..8801539H, 2021Galax...9...58G}. These collimated outflows propagate on extragalactic scales and carry energy and angular momentum away from the central source. They abruptly terminate their journey as bright radio hotspots, where the jet shocks the ambient medium and inflates a giant radio bubble downstream, also known as the lobe. The relativistic speeds of the flow combined with the typical size and magnetic field strength involved make extragalactic jets and their lobes some of the few astrophysical objects capable of confining the most energetic cosmic rays received on Earth \citep{1984ARA&A..22..425H, 2010MNRAS.408L..46G, 2018MNRAS.479L..76M}, but how these particles are accelerated to such ultra-high energies is still an open question.

The jet termination shock itself is a prime candidate for the location of particle acceleration. Unfortunately, it is well established that relativistic uniform shocks are poor accelerators because the transverse nature of the magnetic field with respect to the shock normal drastically reduces the efficiency of diffusive shock acceleration (\citealt{1990ApJ...353...66B, 2018MNRAS.473.2364B}, see however \citealt{2023MNRAS.519.1022K}). Even under the most favorable conditions of an unmagnetized shock, particle acceleration via this mechanism would be too slow to explain the extreme energization rate required to reach the confinement limit of the source \citep{2013ApJ...771...54S, 2018MNRAS.477.5238P}. In contrast, relativistic magnetic reconnection is sufficiently fast to alleviate the above difficulties, but particle acceleration up to the highest energies would require a system-size current sheet and an unrealistically high plasma magnetization \citep{2016ApJ...816L...8W}. A large-scale velocity shear provides another way to accelerate particles efficiently \citep{1998A&A...335..134O, 2022Univ....8..607R}. This situation could be found for instance at the interface between the jet and its backflow or the external medium (e.g., \citealt{2014NJPh...16c5007A, 2018PhRvD..97b3026K, 2019MNRAS.482.4303M, 2021ApJ...907L..44S, 2023MNRAS.519.5410M, 2023MNRAS.519.1872W}), but it is unclear how it could operate at the jet hotspots and lobes where in situ particle acceleration must occur.

In this work, we present a novel ab initio plasma model of particle acceleration at the jet termination shock beyond the local uniform field approximation that has been used so far by considering the full radial dependence of the magnetic field across the jet. For this purpose, a new numerical setup inspired by a force-free jet solution is introduced in Sect.~\ref{sect::setup}. Sect.~\ref{sect::results} focuses on the dynamics of the flow, acceleration of the ions, and their escape in the downstream medium. Sect.~\ref{sect::ccl} summarizes the main findings of this work and their astrophysical implications.

\section{Numerical methods and setup}\label{sect::setup}

We focus our attention on the global structure of the shock front and nonthermal particle acceleration using particle-in-cell simulations performed with the {\tt Zeltron} code \citep{2013ApJ...770..147C}. The plasma is so dilute that it can be considered as collisionless. The electron-ion plasma dynamics was simulated with an artificially low mass ratio $m_{\rm i}/m_{\rm e}=25$ to reduce the computational cost, but it has remained sufficiently large to preserve a safe scale separation \citep{2013ApJ...771...54S}. The jet was modeled as a cylindrical plasma column of radius $R_{\rm j}$ moving with a uniform bulk Lorentz factor $\Gamma_{\rm j}=1/(1-V^2_{\rm j}/c^2)^{1/2}=10$ along the $+z$ direction. The exact magnetic structure in the jet is not well constrained by observations, but it is most likely dominated by the toroidal component far away from the jet-launching point \citep{2021Galax...9...58G}. This component stems from the twist of poloidal field lines by the accretion disk orbiting around the central black hole \citep{1982MNRAS.199..883B} or by the spin of the black hole itself \citep{1977MNRAS.179..433B}.

\begin{figure}
\centering
\includegraphics[width=\hsize]{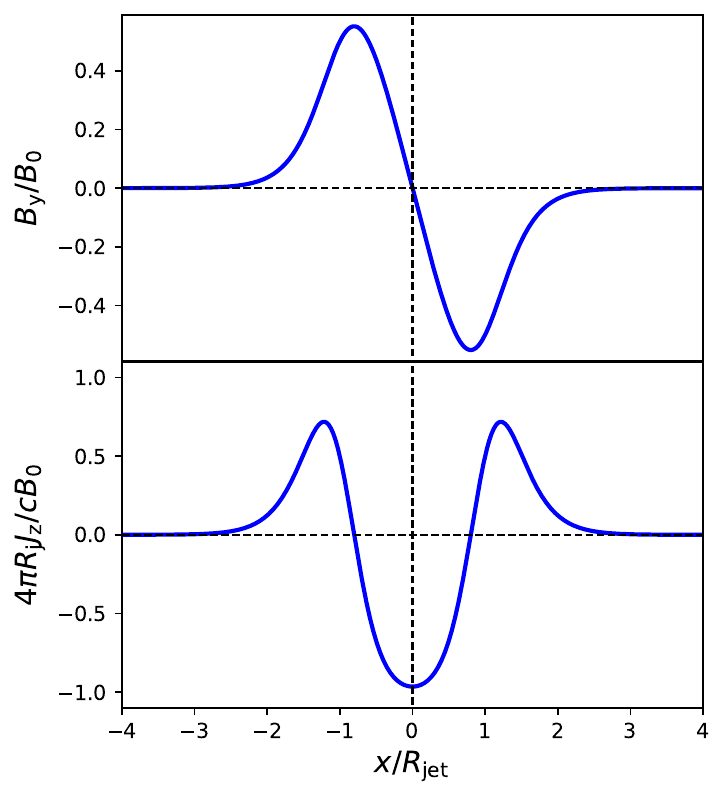}
\caption{Transverse magnetic field strength (top panel, $B_y$) and current density (bottom panel, $J_z$) profiles carried by the jet in the upstream medium projected into the $xz$ plane, where $\Delta=0.5 R_{\rm j}$.}
\label{fig:fig1}
\end{figure}

For a steady, axisymmetric, and Poynting-flux-dominated jet at launching, the toroidal component is obtained from the relativistic Grad-Shafranov equation \citep{1973ApJ...180L.133M, 1973ApJ...182..951S}. Assuming an initial uniform poloidal magnetic field, $B_{\rm p}$, spinning at a constant angular velocity $\Omega$ within the jet radius and gradually cutting off outside, such that $\boldsymbol{\Omega}\cdot\mathbf{B_p}>0$, the solution is
\begin{equation}
B_{\phi}=-\frac{B_0}{2}\left(\frac{R}{R_{\rm j}}\right)\left[1-\tanh\left(\frac{R-R_{\rm j}}{\Delta}\right)\right],
\end{equation}
where $B_0$ is the fiducial magnetic field strength, $R$ is the cylindrical radius to the jet axis, and $\Delta=0.5 R_{\rm j}$ is fixed throughout this study. The hyperbolic tangent term has been added here to parametrize a smooth and finite physical edge of the jet within the simulation domain. The jet must transport a current density $\mathbf{J}=c\boldsymbol{\nabla}\times\mathbf{B}/4\pi$, whose nonvanishing component is along the jet axis,
\begin{equation}
J_{z}=-\frac{cB_0}{4\pi R_{\rm j}}\left[1-\tanh\left(\frac{R-R_{\rm j}}{\Delta}\right)-\frac{R}{2\Delta}\cosh^{-2}\left(\frac{R-R_{\rm j}}{\Delta}\right)\right].
\end{equation}
The first two terms represent a net negative current density uniformly distributed within the core of the jet, while the last term is the positive (return) current that flows along the jet boundary and satisfies the current closure in the system, $\boldsymbol{\nabla}\cdot\mathbf{J}=0$, akin to a coaxial cable (see, Fig.~\ref{fig:fig1}). The relativistic bulk motion of the plasma along the jet direction is accompanied by the ideal motion electric field $\mathbf{E}=-\mathbf{V_{\rm j}}\times\mathbf{B}/c$, whose nonvanishing component is along the radial direction. Thus, the plasma is polarized: the core of the jet is negatively charged, while the sheath is positively charged (and vice versa if $\boldsymbol{\Omega}\cdot\mathbf{B_p}<0$). The motion of this net charge density $\rho$ carries the required current. Although polarized, the plasma remains quasi-neutral: $\left|\rho\right|/e\ll n_0$, where $e$ is the electron charge and $n_0$ is the reference plasma density in the jet.

The simulation box is limited to a two-dimensional Cartesian plane ($xz$ coordinates) mimicking the poloidal plane of a cylindrical jet ($Rz$ coordinates). With this choice of coordinates, the expansion of the flow with distance from the jet axis and the drift motion of the plasma associated with the curvature of field lines are missing \citep{2023arXiv230408132H}. Yet, the Cartesian geometry has the advantage of capturing some of the nonaxisymmetric features that would be missing in a two-dimensional cylindrical simulation. The departure from perfect axisymmetry plays an important role in the escape mechanism of energetic particles as shown below. We defer the study of full three-dimensional simulations of a cylindrical jet to a future work.

The fiducial run is composed of $16,384\times262,144$ cells along the $x$ and $z$ directions, respectively. The box is elongated along the jet direction, it extends between $-4R_{\rm j}<x<-4R_{\rm j}$ and $0<z<128 R_{\rm j}$. The reference electron plasma skindepth, $d_{\rm e}=(\Gamma_{\rm j}m_{\rm e}c^2/4\pi n_0 e^2)^{1/2}$, has been resolved by  eight cells and the ion skindepth scale $d_{\rm i}=d_{\rm e}(m_{\rm i}/m_{\rm e})^{1/2}$ by $40$~cells, meaning that the box size is $410 d_{\rm i}\times6554 d_{\rm i}$. The simulation time step has been fixed at $\omega_{\rm pe}\Delta t=8.75\times 10^{-2}$, where $\omega_{\rm pe}=c/d_{\rm e}$ is the electronic plasma frequency. All boundaries are perfectly reflective for both the fields and the particles. The $z=z_{\rm max}$ boundary mimics the contact discontinuity with the ambient medium that is not simulated here to save on computational resources. Another common optimization procedure is to inject the plasma into the box gradually. Initially, the particle injector is close to the $z_{\rm max}$ wall, and then it recedes away from this boundary at the speed of light. The injected plasma bounces off the wall with no loss of energy such that the beam interacts with itself and the shock forms. The choice of reflecting boundaries on the $\pm x$ directions mimics the presence of a perfectly confining medium surrounding the jet. At injection, the plasma was modeled by eight~particles in each cell with a constant density at all radii, including the region outside the jet boundaries. The particles are streaming at the jet bulk Lorentz factor, but a small temperature $T/m_{\rm e,i}c^2=0.01$ has been included in addition to spatial filtering of the current densities to limit the growth of spurious Cherenkov radiation \citep{2004JCoPh.201..665G}.

An initially Poynting-flux-dominated jet gradually converts its magnetic energy into plasma bulk motion, such that if internal dissipation along its way to the termination shock is weak as in Fanaroff-Riley class II jets \citep{1974MNRAS.167P..31F}, the magnetic field and the particles could be close to equipartition \citep{1992ApJ...394..459L, 2007MNRAS.380...51K}. In this work, we assume perfect equipartition which translates into the magnetic enthalpy density to the particle enthalpy density ratio, $\sigma\equiv B_0^2/4\pi \Gamma_{\rm j} n_0 m_{\rm i} c^2=1$ (the electronic contribution is neglected in this expression due to the mass ratio difference). We also ran the same numerical setup with $\sigma=0.1$ and found similar results. We focus this study on the case where the poloidal magnetic component is negligible, $B_{z}=0$. In reality, it is likely that a small but finite poloidal component still exists and stabilizes the jet against current driven instabilities such as the kink mode which could lead to its disruption akin to a Z-pinch configuration in full 3D (e.g., \citealt{1954RSPSA.223..348K, 1998ApJ...493..291B, 1999MNRAS.308.1006L, 2010MNRAS.402....7M}). We performed two additional simulations with $B_{z}/B_0=0.01$ and $B_{z}/B_0=0.1$, but we did not find significant differences with respect to the fiducial run presented below. Table~\ref{table:sim} summarizes all the runs performed in this work.

\begin{table}
\caption{\label{table:sim}List of all the simulations reported in this work.}
\centering
\begin{tabular}{lcccc}
\hline\hline
Run & Cells & Size (in $d_{\rm i}$) & $\sigma$ & $B_z/B_0$\\
\hline
Fiducial & $16,384\times262,144$ &  $410\times6554$   & $1$ & $0$ \\
S01      & $8,192\times65,536$   &  $205\times1638$   & $0.1$ & $0$ \\
Bz0      & $8,192\times65,536$   &  $205\times1638$   & $1$ & $0$ \\
Bz001    & $8,192\times65,536$   &  $205\times1638$   & $1$ & $0.01$ \\
Bz01     & $8,192\times65,536$   &  $205\times1638$   & $1$ & $0.1$ \\
\hline
\end{tabular}
\end{table}

\section{Results}\label{sect::results}

\subsection{Shock structure and evolution}

\begin{figure}
\centering
\includegraphics[width=\hsize]{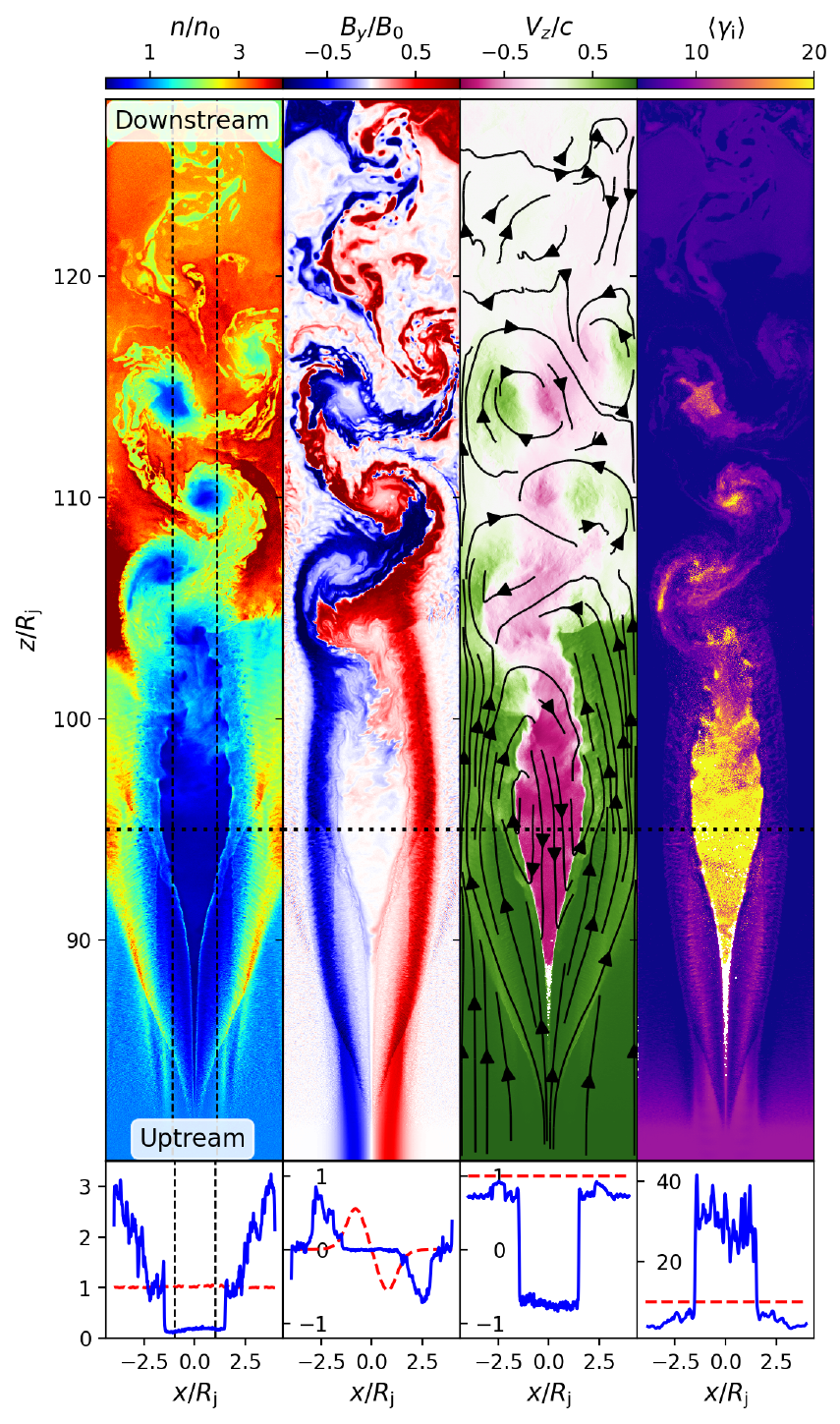}
\caption{Structure of the jet termination shock in the final state of the fiducial simulation at $\omega_{\rm pi}t=2500$. From left to right, this figure shows the following: the plasma density $n/n_0$, the out-of-plane magnetic field component ($B_y/B_0$), the axial plasma bulk velocity ($V_{z}/c$) and streamlines (solid contour lines with arrows), as well as the mean individual ion Lorentz factor $\langle\gamma_{\rm i}\rangle$. The lower panels show the radial profile of each quantity across the large plasma cavity at $z\approx 95 R_{\rm j}$ (blue solid line), which can be compared with the upstream solution (red dashed line).}
\label{fig:fig2}
\end{figure}

\begin{figure}
\centering
\includegraphics[width=\hsize]{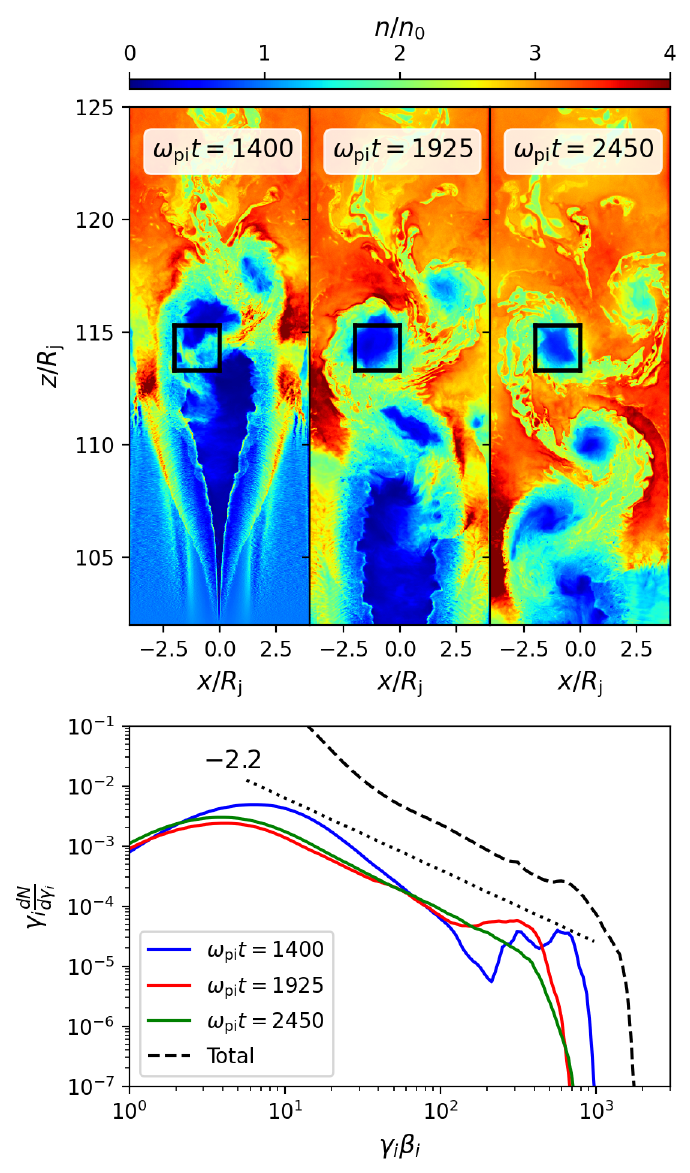}
\caption{Zoomed-in view and time evolution of the von~K\'arm\'an vortex street. The top panels shows the formation of a vortex that was carved out of the shock front cavity, and its later evolution in the downstream medium. The bottom panel shows the evolution of the ion spectrum within the vortex encapsulated in the box drawn in the upper panels. The total final ion spectrum is shown for comparison with the black dashed line.}
\label{fig:fig3}
\end{figure}

\begin{figure}
\centering
\includegraphics[width=\hsize]{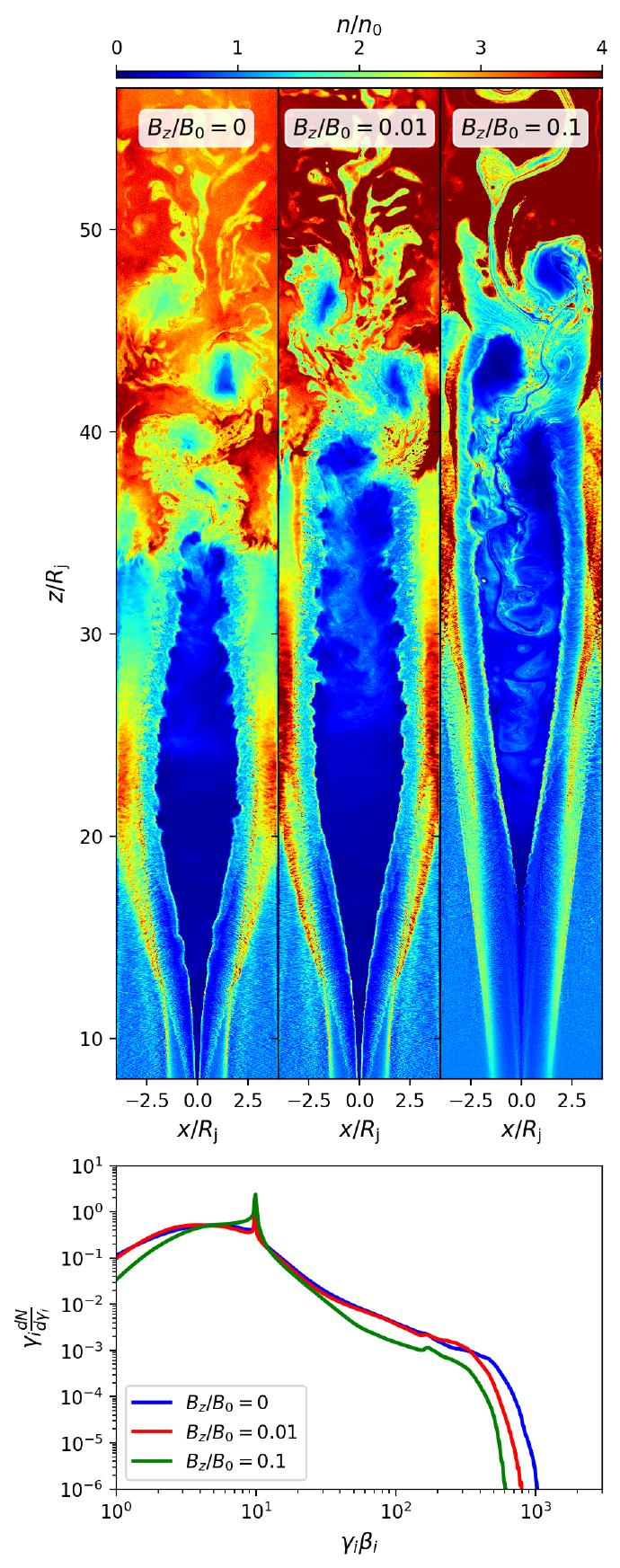}
\caption{Termination shock structure (top panels: plasma density) and total ion spectrum (bottom panel) as a function of the strength of the vertical magnetic field component $B_{z}/B_0=0,$ $0.01$ and $0.1$ at $\omega_{\rm pi} t=1575$.}
\label{fig:fig4}
\end{figure}

Fig.~\ref{fig:fig2} shows the final state of the fiducial simulation. After the initial reflection of the beam, the flow quickly reconfigured and produced a shock and the downstream medium. For $\left|x\right|>R_{\rm j}$ where $\sigma\approx 0$, a Weibel-mediated shock formed, which is in agreement with previous studies of unmagnetized relativistic shocks \citep{2003ApJ...595..555N, 2013ApJ...771...54S}. This solution is characterized by the growth of multiple thin plasma filaments near the shock, in both the upstream and the downstream media, generated by particles reflected at the shock front. Within the jet radius, the flow behaves very differently. The plasma collapses on the axis where the magnetic field vanishes due to the partial loss of the support from the jet electric force. Near the edges, the shock becomes gradually more magnetized ($\sigma\sim 1$) and therefore weaker \citep{1984ApJ...283..694K}. The most striking feature is the formation of an over-pressured cavity at the shock front along the jet axis that is drilling through the upstream medium. A similar pattern has recently been reported in the context of pulsar-wind termination shocks in pair plasmas \citep{2020A&A...642A.123C}, but this phenomenon differs in this work by its extreme magnitude. The lateral size of this cavity does indeed increase with time up to the point where it fills the full radial extent of the jet. The cavity is under-dense ($\sim 0.1$-$0.2 n_0$), weakly magnetized, and filled with energetic particles (see Fig.~\ref{fig:fig2}). At this stage, the Weibel-mediated shock that prevailed in the early phases of the simulation has nearly completely disappeared because of the overwhelming presence of the cavity. In fact, it becomes even difficult to define a clear shock front in the simulation. It can be approximately located at the base of the cavity where the flow significantly decelerates and the plasma density increases by a factor $\approx 3$ at $z\approx 105 R_{\rm j}$ in Fig.~\ref{fig:fig2}.

Another important property is the presence of a strong backflow in the cavity moving relativistically toward the upstream medium against the jet at a velocity of $V_{z}\approx -0.75c$. A sharp tangential velocity discontinuity between the edges of the cavity and the incoming flow drives the formation of strong Kelvin-Helmholtz vortices with a size on the order of the width of the cavity itself at any given time. The downstream medium close to the shock front is dominated by the dynamics of an oscillating wake trailing behind the cavity. The wake periodically carves part of the cavity out, which is then advected further downstream in the form of a series of hot, low-density bubbles (Fig.~\ref{fig:fig2}), resembling a von~K\'arm\'an vortex street. Their position is nearly at rest with respect to the simulation frame (see top panels in Fig.~\ref{fig:fig3}), but the flow inside the bubbles has a strong velocity vorticity which in turn leads to efficent mixing and its progressive obliteration in the downstream medium. We identify this phenomenon as the main escape mechanism which allows energetic particles momentarily stored in the cavity to move away from the acceleration zone. The shock front cavity and vortices also form in the presence of a weak vertical magnetic field component, $B_{z}/B_0=0.01$, $0.1$; we observe a stronger plasma compression in the downstream medium with increasing $B_z$ (top panels in Fig.~\ref{fig:fig4}).

\subsection{Particle acceleration}

\begin{figure}
\centering
\includegraphics[width=\hsize]{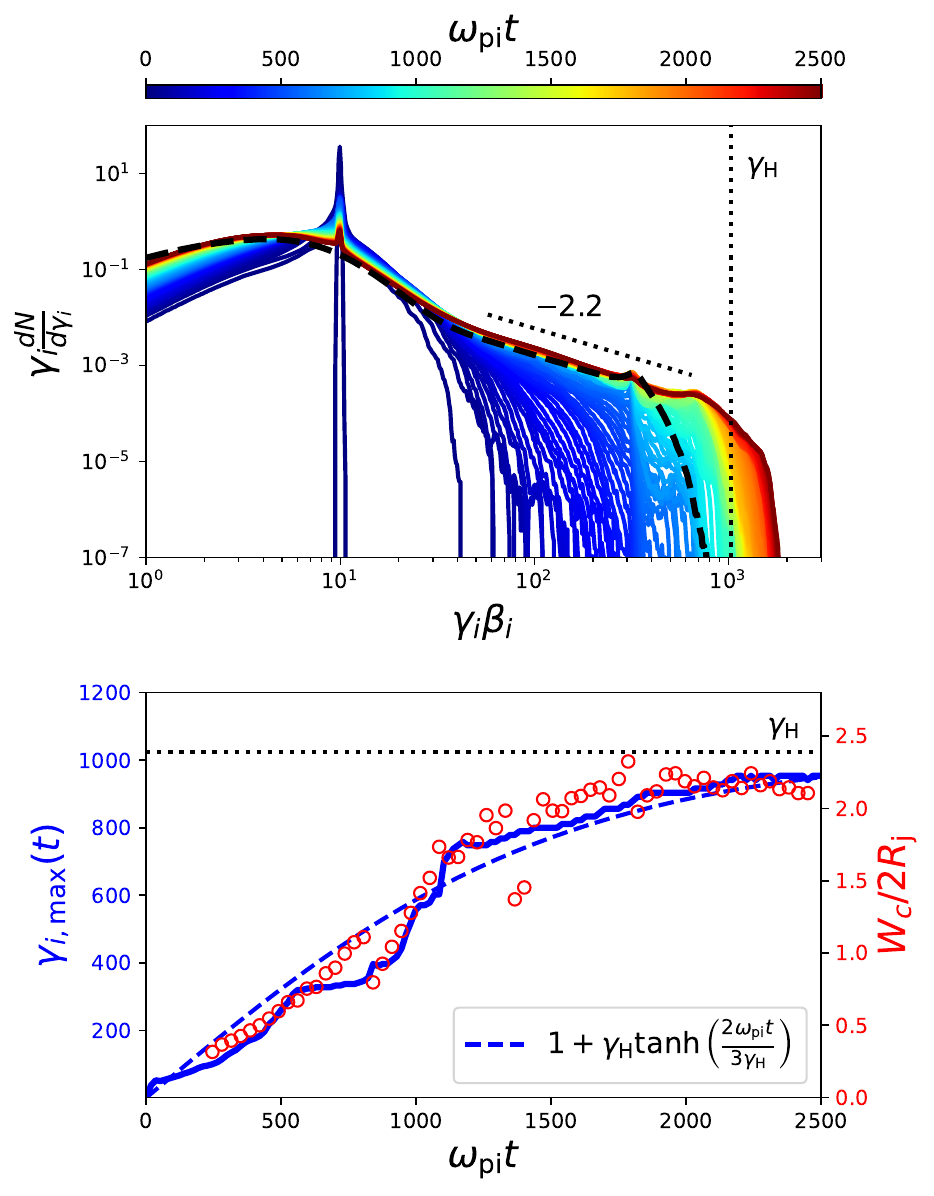}
\caption{Particle acceleration up to the confinement limit of the system. Top panel: Time evolution of the total ion spectrum $\gamma_{\rm i} dN/d\gamma_{\rm i}$ (color-coded solid lines). The final total electron spectrum is shown for comparison as a function of $\gamma_{\rm e}\beta_{\rm e}\times(m_{\rm e}/m_{\rm i})$ to compare the energy scales rather than the Lorentz factors between the two species (black dashed line). A $-2.2$ power-law slope is shown for comparison (dotted line). Bottom panel: Time evolution of the maximum ion Lorentz factor $\gamma_{\rm i,max}$ (blue solid line) and the latitudinal width of the shock front cavity normalized to the full jet width ($W_c$, red circles). The jet confinement limit, $\gamma_{\rm H}$, is marked by the horizontal dotted line. The blue dashed line shows a simple empirical formula that captures the time evolution of $\gamma_{\rm i,max}$.}
\label{fig:fig5}
\end{figure}

\begin{figure}
\centering
\includegraphics[width=\hsize]{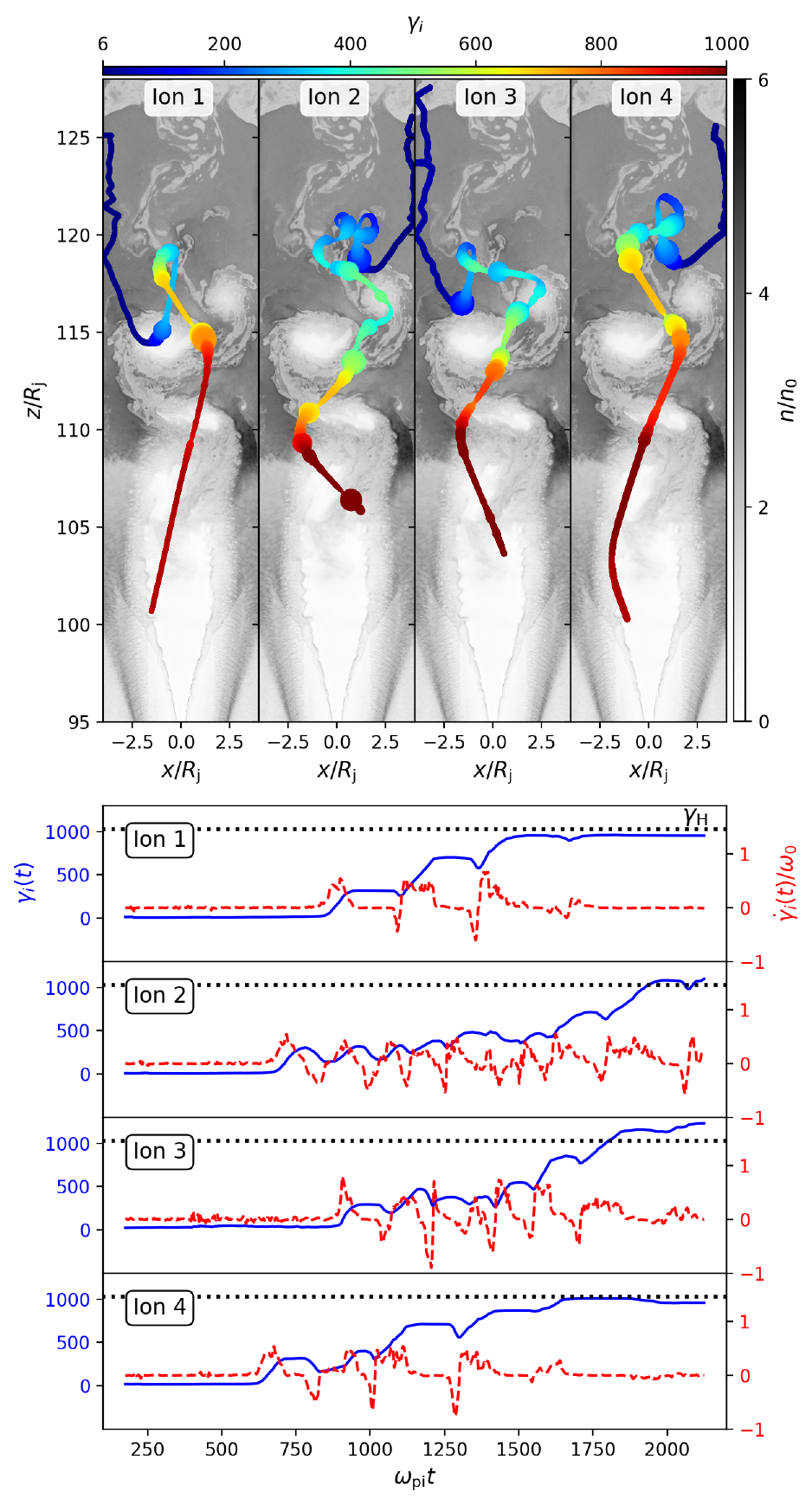}
\caption{Acceleration history of four representative high-energy ion trajectories tracked from their injection at $\omega_{\rm pi}t=175$ up to $\omega_{\rm pi}t\gtrsim 2000$ where they have all reached the jet confinement limit. Top: Full ion trajectory superimposed onto the final plasma density map (gray scale). The color indicates the particle Lorentz factor $\gamma_{\rm i}$, while the marker size codes for the acceleration rate $\dot{\gamma}_{\rm i}/\omega_{0}$, where $\omega_0=eB_0/m_{\rm i}c$, a large marker means a high positive rate while a small marker is for negative rates. Bottom: Time evolution of the ion Lorentz factor (blue solid line) and acceleration rate (red dashed line).}
\label{fig:fig6}
\end{figure}

In contrast to plane-parallel uniform magnetized shocks, the strong variations of the transverse magnetic field strength and, in particular, the presence of a magnetic null line along the jet axis leads to an extremely efficient acceleration of particles. Fig.~\ref{fig:fig5} shows that the total particle spectrum is composed of a low-energy thermal bump centered around the upstream jet Lorentz factor $\Gamma_{\rm j}$, and a high-energy power-law tail with an index $\approx -2.2$ extending up to the accelerator confinement limit defined by the Hillas criterion \citep{1984ARA&A..22..425H},
\begin{equation}
\gamma_{\rm H}\equiv \frac{2 R_{\rm j}eB_0}{m_{\rm i}c^2} \approx 1000,
\end{equation}
in the fiducial simulation. A clear sign that the limit of the accelerator has been reached is the saturation of the maximum ion spectrum cut-off Lorentz factor, $\gamma_{\rm i,max}$, as it approaches $\gamma_{\rm H}$. This saturation is visible as a small bump in the spectrum just below the high-energy cutoff. In the early phases, $\gamma_{\rm i,max}$ grows approximately linearly with time, which is compatible with an ideal Bohm-like acceleration regime. The full evolution is well approximated by the following empirical formula
\begin{equation}
\gamma_{\rm i,max}\left(t\right)\approx 1+\gamma_{\rm H}\tanh\left(\frac{2\omega_{\rm pi}t}{3\gamma_{\rm H}}\right).
\end{equation}
Fig.~\ref{fig:fig5} also demonstrates the perfect coevolution of the maximum particle energy with the width of the cavity, establishing an unambiguous connection between particle acceleration and the growth of the cavity whose dynamics is governed by the cosmic-ray pressure. The final electron spectrum has a very similar shape as the ion spectrum, but it cuts off at slightly lower energies, $\gamma_{\rm e,max}\times\left(m_{\rm e}/m_{\rm i}\right)\approx 300$. The asymmetry between high-energy electrons and ions is explained by the slightly different acceleration patterns followed by each species and may depend on the polarization of the jet (i.e., the sign of $\mathbf{B_p}\cdot\boldsymbol{\Omega}$, \citealt{2018ApJ...863...18G, 2020A&A...642A.123C}). This effect may also be weaker at higher scale separation. We observe a decrease in the particle acceleration efficiency with increasing strength of the vertical field component (Fig.~\ref{fig:fig4}, bottom panel).

To elucidate the origin of this extreme particle acceleration, we tracked a sample of $20,000$ simulation particles for each species. They have been randomly chosen throughout the simulation box to avoid any selection biases. The high-energy ions that reach the confinement limit all follow a very similar acceleration history as shown in Fig.~\ref{fig:fig6}. The time evolution of the ion energy is characterized by discrete and intense acceleration episodes during which the particle Lorentz factor grows almost linearly. Far from the saturation, the energy gain during each step is close to $\Delta\gamma\sim \gamma$. The accelerate rate $\dot{\gamma}$ is also independent of the particle energy. We find that each of these acceleration events are unambiguously related to the crossing of velocity shear layers. The first energy boost occurs when the particle crosses the main shear layer at the interface between the upstream medium and the shock front cavity. Once they are captured by the cavity, they are further accelerated in the wake trailing behind. The coherent oscillating motion of the wake allows for multiple crossings of the velocity shear layer. An acceleration episode is often preceded by a brief deceleration event associated with the sudden large-amplitude change in the direction of the momentum that the particle experiences as it encounters an oppositely directed flow.

The particles shown here reach an energy saturation and remain in the cavity until the end of the simulation without significant further energization. Looking at other high-energy particles shows that, eventually, they are captured in a bubble and advected in the downstream medium without a significant loss of energy, which allows them to escape the acceleration site. In fact, the spectrum measured within a von~K\'arm\'an vortex presents two components soon after its formation: a low-energy bump and a narrow high-energy component cutting off near the confinement limit. At later times when the vortex is further downstream, the spectrum evolves to a single power-law distribution with an index $\sim -2.2$ without significant changes in the high-energy cutoff (see bottom panels in Fig.~\ref{fig:fig3}).

Particles are accelerated by experiencing the large-scale ideal electric field associated with the varying bulk motion of the flow across the velocity shear layers, such that the acceleration rate is $\dot{\gamma}\sim e\mathbf{E}\cdot\mathbf{V}/m_{\rm i}c^2\lesssim 0.5 \omega_0$, where $\omega_0=eB_0/m_{\rm i}c$. This explains the high acceleration efficiency, and is in stark contrast with the microscopic turbulent nature of the scattering centers that is expected in relativistic Weibel-dominated shocks. Particle acceleration at a shear layer can be understood as a Lorentz boost of the particle energy as it is scattered by plasma turbulence frozen into the flow approximately moving at the $\mathbf{E}\times\mathbf{B}$-drift velocity. For an isotropic particle distribution near the shear layer, the mean energy gain per crossing is $\langle\Delta\gamma/\gamma\rangle=\Gamma_{\rm s}-1$, where $\Gamma_{\rm s}=1/(1-V^2_{\rm s}/c^2)^{1/2}$ and $V_{\rm s}$ is the relative velocity across the layer \citep{1990A&A...238..435O, 2004ApJ...617..155R}. With typical flow velocities $V_{\pm}\sim\pm 0.5c$ across the boundary gives $V_{\rm s}=(V_+-V_-)/(1-V_+ V_-/c^2)\sim 0.8c$, so a net energy gain per crossing $\langle\Delta\gamma/\gamma\rangle\sim 1$ in agreement with our findings. The $\approx -2$ slope of the nonthermal power-law tail in the particle spectrum can also be well accounted for by a simple Fokker-Planck model assuming a particle scattering time on the order of the Larmor gyration time \citep{1998A&A...335..134O, 2006ApJ...652.1044R}.

\section{Conclusion}\label{sect::ccl}

The toroidal nature of the magnetic field within the jet leads to the formation of an over-pressured cavity filled with cosmic rays and a relativistic shear flow localized near the shock front. These features play a key role in the extreme particle acceleration mechanism reported in this work. Particles are primarily accelerated at the boundaries of the cavity, where the tangential velocity discontinuity is strongest, up to the confinement limit of the system. The cavity acts as an obstacle to the incoming flow and generates a strong von~K\'arm\'an vortex street-like structure that allows energetic cosmic rays to escape in the downstream medium. The magnetic field inside the cavity is weak such that the high-energy particles it contains should not radiate much via synchrotron emission, in contrast to the downstream medium. Therefore, the cavity may be observable as a dark region in the vicinity of the jet hotspots. Interestingly, a hole in the radio and X-ray images of Cygnus~A jets has been recently reported \citep{2020ApJ...891..173S} and, possibly, also in Pictor~A jets \citep{2022ApJ...941..204T}, it would be worthwhile to check whether this feature is ubiquitous among other similar jets as predicted by the model.

Scaling our results up to the typical physical conditions of extragalactic jet hotspots and lobes, assuming a $B_0\sim 10\mu$G, a $L\sim 10$kpc size yields a maximum proton energy at saturation on the order of $\mathcal{E}_{\rm p,max}=\gamma_{\rm p,max}m_{\rm p}c^2\sim eB_0 L\sim 10^{20}$eV. Thus, the findings of this work support the idea that not only are extragalactic jets able to confine ultra-high-energy cosmic rays, but they may also be capable of generating them. The configuration explored in this study is generic, and therefore the conclusions are applicable to other astrophysical systems involving an organized large-scale toroidal magnetic field. In another work, we argue that a similar mechanism could operate in relativistic pair plasmas at the shock front of pulsar wind nebulae, and explain extreme particle acceleration up to PeV energies \citep{2020A&A...642A.123C}. The termination shock of stellar-mass black hole jets such as in the microquasar SS433 \citep{2004ASPRv..12....1F} is another relevant astrophysical environment where this mechanism could take place. The recent discovery of $>10$~TeV gamma rays from the lobes spatially coincident with compact X-ray knots suggests that extreme in situ particle acceleration close to PeV energies must take place \citep{2018Natur.562...82A}, which is an outstanding challenge that may be overcome by the mechanism highlighted in this work.

\begin{acknowledgements}
We thank B.~Crinquand and M.~Ostrowski for their comments on an early draft, and the referee Luke~Drury for his comments and supportive report. This project has received funding from the European Research Council (ERC) under the European Union’s Horizon 2020 research and innovation program (Grant Agreement No. 863412). Computing resources were provided by TGCC and CINES under the allocation A0130407669 made by GENCI.
\end{acknowledgements}

\bibliographystyle{aa}
\bibliography{jet_ts}

\end{document}